\documentclass[twocolumn,floatfix]{aastex63}

\usepackage{amssymb}
\usepackage{amsmath}
\usepackage{hyperref}
\usepackage{verbatim}
\usepackage{float}

\usepackage{graphicx}

\begin{document}

\title{Do Red Galaxies Form More Stars Than Blue Galaxies?}

\author[0000-0003-3780-6801]{Charles L. Steinhardt}
\affiliation{Department of Physics \& Astronomy, University of Missouri, 701 S. College Ave., Columbia, MO 65201}
\affiliation{Cosmic Dawn Center (DAWN)}
\affiliation{Niels Bohr Institute, University of Copenhagen, Jagtvej 120, DK-2100 Copenhagen \O}


\begin{abstract}
A new model is proposed in which typical galaxies form most of their stellar mass in a phase with an intrinsically red stellar population.  In the standard picture, galaxies with intrinsically red stellar populations are believed to have old stellar populations, so that only galaxies with blue stellar populations have significant star formation, and subsequent changes to the stellar population come from predominantly from aging and merging populations which have already formed.  However, several observational puzzles have developed which are difficult to reconcile with this standard scenario.  The most massive blue star-forming galaxies, presumed to be at the end of their stellar mass growth, are $\sim 1$ dex less massive, have a $\sim 1$ dex lower $M_*/M_{BH}$ ratio, and have a bottom-lighter IMF than local quiescent galaxies.  Here, a new solution is proposed: at low temperature and high metallicity, galaxies can continue to form stars efficiently without being able to form O and B stars.  These red star-forming galaxies would have many of the same properties of the population currently described as post-starburst galaxies, allowing a new interpretation of their origin.  Finally, additional falsifiable observational predictions of this model are also discussed.
\end{abstract}


\section{Introduction}
\label{sec:intro}

The color of a stellar population has long been used as an indicator of whether there is significant ongoing star formation.  The luminosity of a main sequence star scales approximately as $L \propto M^{3.5}$, so that the starlight from a galaxy is dominated by its most massive individual stars, which are also both the bluest and shortest-lived stars.  Thus, the starlight from a galaxy with a young stellar population will appear blue.  When a galaxy stops forming stars, these blue stars die out, and the remaining starlight will instead be red.  As a result, blue galaxies are also often described as star-forming, while many red galaxies are considered quiescent\footnote{This separation is complicated by reddening from dust, which can be formed along with young stars.  Thus, a series of color-color selections have been proposed \citep{Williams2009,AntwiDanso2023} to separate dusty star-forming galaxies with intrinsically blue stellar populations from quiescent ones with red stellar populations.}, due to the inferred lack of recent star formation.  The transition from blue to red occurs rapidly, with relatively few galaxies observed to lie in the `green valley' between blue and red.


Thus, the canonical picture is fairly straightforward.  Following an initial assembly period, galaxies are able to form stars efficiently, and during this period they comprise the identified population of star-forming galaxies.  At some point, via one of many proposed potential quenching mechanisms, these processes cease.  After a few hundred Myr, the color changes from blue to red and galaxies can be identified as quiescent.  Once a galaxy is fully quenched, its quenching is believed to be permanent and to mark the end of efficient star formation, although recent studies have found a small population of galaxies which might be `rejuvenated' \citep{Zhang2023}.  With no additional significant star formation or quasar accretion past this point, any further changes to the stellar population can only come from aging and merging populations which have already formed.

Here, several significant differences between blue star-forming stellar populations and those measured in local quiescent galaxies (summarized in Table \ref{tab:differences}) are presented which cannot be explained by aging and merging alone.  In order to explain these differences, a new scenario is proposed in which most of the final stellar mass in a galaxy is instead produced in a phase with an intrinsically red stellar population which would have similar spectra to post-starburst galaxies.  An astrophysical mechanism to allow red galaxies to form stars efficiently and potential tests for whether these red star-forming galaxies comprise a subset of the current post-starburst population are also described.  

In \S~\ref{sec:puzzles}, several observations are described which are difficult to reconcile with the standard quenching scenario when comparing high-redshift galaxies near the end of their star formation with the local quiescent population.  Specifically, this scenario appears to be incompatible with the black hole mass - bulge mass relation (\S~\ref{subsec:magorrian}) and the presence of the most massive local galaxies (\S~\ref{subsec:m87}) as well as with observations of a bottom-heavy IMF in local quiescent galaxies (\S~\ref{subsec:bottomheavy}), and with observations of `post-starburst' galaxies.  In \S~\ref{sec:rsfg}, a new model is proposed in which red galaxies continue to form stars after the point at which they appear to quench, but with a low Jeans mass and a bottom heavy-IMF.  Galaxies in this phase would have several properties in common with the significant observed population of post-starburst galaxies, allowing a reinterpretation as red star-forming galaxies (\S~\ref{subsec:poststarburst}).  This model is shown to potentially solve both problems in \S~\ref{sec:solution}.  Finally, the implications of this scenario and additional falsifiable predictions are described in \S~\ref{sec:discussion}.

\section{Observations Difficult to Reconcile with the Canonical Scenario}
\label{sec:puzzles}

Under the canonical paradigm for galaxy evolution, stellar mass growth is associated with star-forming galaxies, or more specifically, with periods when galaxies contains young stellar populations with their light dominated by short-lived O and B stars.  Any further changes to the stellar population must come through aging and merging rather than from additional star formation.  Although this is intuitive and perhaps even obvious, there are several significant differences between blue star-forming galaxies at turnoff and local quiescent galaxies (Table \ref{tab:differences}).  Here, two observational results are described which appear inconsistent with this picture.  
\begin{table*}[!ht]
  \centering
  \begin{tabular}{|c|c|c|c|c|c|c|}
    \hline
    \textbf{Property} &\textbf{Blue SFG} & \textbf{Quiescent Galaxies} & \textbf{Effect of Aging} & \textbf{Effect of Merging} & \textbf{Proposed Red SFG} \\
    \hline
    Color & Blue & Red & Becomes redder & None & Becomes red \\
    \hline
    Max $M_*$ & $10^{11.5} M_\odot$ & $ > 10^{12.3} M_\odot$ & Negligible & Increase along with $M_{BH}$ & $\sim 1$ dex increase \\
    \hline
    Max $M_{BH}$ & $10^{9.8} M_\odot$ & $10^{9.8} M_\odot$ & Negligible & Increase along with $M_*$ & Negligible \\
    \hline
    $M_* / M_{BH}$ & $\sim 30$ & $\gtrsim 300$ & Negligible & None & Increase by $\sim 1$ dex \\
    \hline
    Stellar MF & Bottom-light & Bottom-heavy & None & None & Becomes bottom-heavy \\
    \hline
  \end{tabular}
  \caption{Summary of several observed differences between blue star-forming galaxies at higher redshift (SFG) and red quiescent galaxies in the local Universe.  Although aging and merging can change the overall color and can add stellar mass, they cannot change the $M_* / M_{BH}$ ratio or change a bottom-light stellar population to a bottom-heavy one.  An additional red star-forming phase is proposed in \S~\ref{sec:puzzles} as a potential solution, and would be able to explain the missing stellar mass.}
  \label{tab:differences}
\end{table*}

\subsection{Magorrian Relation}
\label{subsec:magorrian}

\begin{figure}
    \centering
    \includegraphics[width=0.45\textwidth]{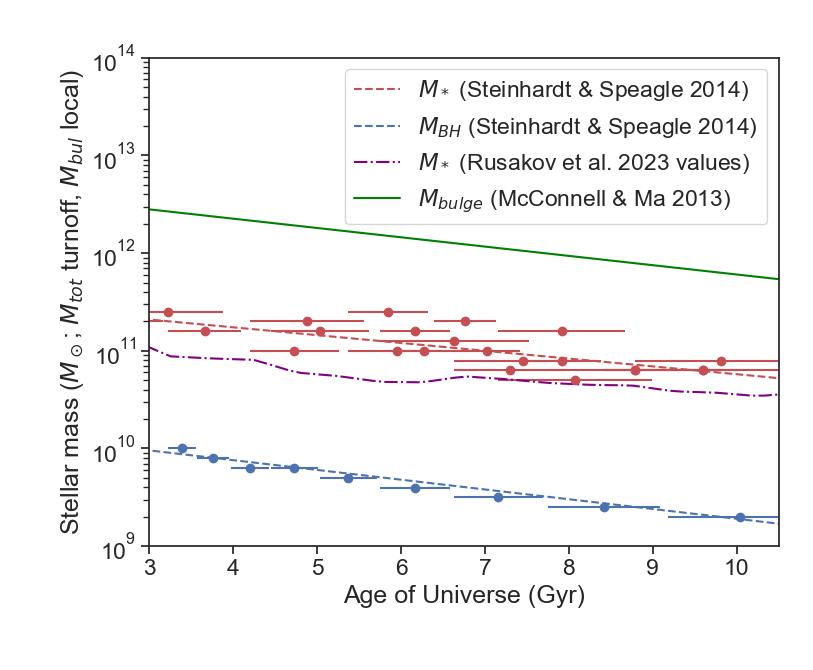}
    \label{fig:downsizing}
    \caption{At turnoff, the ratio $M/_*/M_{BH}$ is approximately 30 \citep{Steinhardt2014b} (red, dashed) assuming that star-formation and quasar accretion cease concurrently.  However, in the local Universe, with the bulge stellar mass alone it is closer to 300:1 \citep{Haring2004,McConnell2013} (blue points; green best-fit relation).  Therefore, at some point after galaxies are no longer blue, the ratio must increase by an additional factor of $\sim 10$ to match local observations.  The required ratio becomes slightly stronger with improved stellar mass estimates that incorporate a variable IMF from the \citet{Rusakov2023} COSMOS2020 catalog}
\end{figure}

Because of downsizing, the most massive star-forming galaxies at any fixed redshift are more massive than star-forming galaxies at later times, and therefore must be about to quench.  Thus, a turnoff mass can be defined as the mass at which the number density of star-forming galaxies at time $t$ is ten times greater 0.5 Gyr later, with the time interval chosen as an approximation of the time a galaxy takes to change from blue to red after quenching.  Using this definition, turnoff masses are calculated using the value-added properties catalog \citep{Rusakov2023} for COSMOS2020 photometry \citep{Weaver2022}.  With a corresponding definition, similar downsizing behavior allows turnoff black hole masses to be found from the DR11 virial mass catalog \citep{Shen2011}.  The resulting comparison shows that both star-forming galaxies and quasars exhibit similar downsizing (Fig. \ref{fig:downsizing}.  At a wide range of fixed redshifts, the ratio of the turnoff $M_*$ to $M_{BH}$ is approximately 30:1 \citep{Steinhardt2014b}.  Similar behavior is found at $z \sim 1$ in additional studies \citep{Farrah2023a,Farrah2023b}.

If the end of star formation indicates the end of stellar mass growth, then the present-day stellar masses of massive galaxies should be similar to those at turnoff.  Similarly, supermassive black hole masses should also be similar to those at turnoff.  Thus, if star-forming galaxies and quasars co-evolve \citep{Kormendy2013}, then $M_* / M_{BH}$ today should also be approximately 30:1.  However observations over a wide mass range instead put 
the {\em bulge} $M_* / M_{BH}$ at 300:1 \citep{Magorrian1998,Haring2004,McConnell2013}.  Using the full galactic stellar mass, the ratio should be even higher.  

Even if there is a timing offset between star formation and quasar accretion, it would not be possible to produce a 300:1 ratio.  Further, supermassive black holes have no known mechanism for losing significant mass within a Hubble time.  And, although virial masses are highly uncertain and difficult to calibrate, quasars near turnoff are typically estimated to lie between 10\% and 50\% of their theoretically maximum Eddington luminosity \citep{Shen2011,Steinhardt2010,Steinhardt2011}.  Thus, it would not be possible for current estimates to overestimate black hole masses by a factor of 10, as would be necessary to produce $M_* / M_{BH} = 300$.  

Thus, if both stellar mass and black hole mass estimates are correct, the remaining possibility is that galaxies continue to grow in stellar mass after the apparent end of their star formation.  Their stellar mass must grow by approximately a factor of 10, so that in the most massive galaxies, most stars are actually formed in this hidden process.

Although galaxies can certainly grow their stellar mass via mergers, the resulting merger of their supermassive black holes will grow black hole mass as well.  The local black hole mass-bulge mass relation holds over approximately 4 dex \citep{Sahu2022}, and every massive galaxy where sensitive studies are possible is observed to contain a supermassive black hole \citep{Kormendy1995}.  The merger of galaxies with $M_* / M_{BH} \sim 30$ will simply result in a larger galaxy with $M_* / M_{BH} \sim 30$\footnote{In principle, because the merged stellar mass would be immediately measured in unresolved galaxy while the supermassive black holes take time to merge, the $M_*/M_{BH}$ ratio might initially increase.  Thus, if at some point it becomes possible to observe $M_*/M_{BH}$ ratios in a large population of galaxies, it will be essential to understand the SMBH merger timescale in order to interpret the results.  However, Galactic and other local observations indicate that there is eventually only one supermassive black hole in each galaxy, so they must eventually merge, again producing a 30:1 ratio.}.  Therefore, what is needed is instead a mechanism to form additional stars {\em without growing the central supermassive black hole and without being selected as a star-forming galaxy}.

\subsubsection{Massive Local Galaxies}
\label{subsec:m87}

Additional evidence for significant post-turnoff stellar mass growth comes from comparing local, massive galaxies with star-forming galaxies at turnoff.  If galaxies do not grow appreciably after turnoff, then quiescent galaxies today should have similar stellar masses to the ones they had at turnoff.  However, instead the most massive local galaxies are far more massive than any known star-forming galaxy.  

For example, M87 is estimated to have a stellar mass of $M_* = 10^{12.3}$ and a black hole mass of $M_{BH} = 10^{9.8}$ \citep{Gebhardt2009,Akiyama2019}.  The black hole mass is similar to the most massive quasars found near turnoff at high redshift.  However, the stellar mass is nearly a factor of 10 larger than the most massive known star-forming galaxies (Fig. 1) even at those high redshifts.  Thus, the stellar mass of M87 must have grown significantly since its apparent turnoff.  

For stellar mass growth alone, merging is typically a plausible pathway.  This is more difficult for the most massive galaxies, since minor mergers only add a small fraction to the existing stellar mass, and the most massive known galaxies are so rare that there are extremely few counterparts for major mergers.  However, simulations do produce M87-like masses via continued merging \citep{Li2018}.  Critically, however, those same mergers would also continue to grow the central supermassive black holes.  Thus, if M87 grows its final factor of $\sim 10$ in stellar mass via mergers, approximately the same should be true for $M_{BH}$.  But, as above, this would retain too small of a ratio between $M_*$ and $M_{BH}$.

An alternative mechanism in which M87 continued to form stars efficiently after its apparent turnoff, without merging and without growing its central supermassive black hole, would provide another solution.  Such a mechanism is proposed in \S~\ref{sec:rsfg}.

\subsection{Bottom-Heavy IMF in Massive Elliptical Galaxies}
\label{subsec:bottomheavy}

The stellar initial mass function (IMF) is essential for measurements of both star formation rates and stellar masses, since stellar light is dominated by the rare, highest-mass stars.  Essentially all measurements of extragalactic stellar masses and star formation rates are in reality only measurements of the most massive stars in their stellar populations, with the vast majority of the mass inferred from observations of the high-mass tail of the distribution.

The Galactic IMF has been notoriously difficult to measure, with several IMFs presently in common use \citep{Salpeter1955,Kroupa2001,Chabrier2003}.  Until recently, it was generally assumed that the Galactic IMF was also a good approximation for the IMF in all star-forming galaxies.  However, because the IMF is predicted to vary with temperature, density, metallicity, and several other properties of star-forming molecular clouds \citep{LyndenBell1976,Larson1985,Dave2008,Chabrier2014}, it should not be universal.  For example, at high redshift, the CMB temperature exceeds that of Galactic star-forming regions, requiring a bottom-lighter IMF \citep{Jermyn2018}. 

\citet{Jermyn2018} used the \citet{Kroupa2001} formalism to argue for an IMF with a dependence upon a single parameter, the minimum fragmentation mass $\widetilde{m}$, below which few stars are seen \citep{Bate2005}.  Although like the Jeans mass, $\widetilde{m}$ should depend upon not only temperature but also density, metallicity, etc., \citet{Jermyn2018} approximate the IMF assuming that the temperature dependence dominates.  Taking $f(T) \equiv \widetilde{m}(T)/m(20\textrm{ K})$ then produces an IMF with
\begin{equation}
\label{eq:imf}
    \xi(m,T) = \frac{dN}{dm} \propto \begin{cases}
    m^{-0.3}, & m < 0.08 M_\odot f(T) \\
    m^{-1.3}, & 0.08 M_\odot f(T) < m < 0.50 M_\odot f(T) \\
    m^{-2.3}, & m > 0.50 M_\odot f(T) \\
    \end{cases}
\end{equation}

Based on this principle, recent studies allowed the gas temperature to be a free parameter, finding a best-fit resulting IMF for galaxies in large photometric catalogs \citep{Sneppen2022,Steinhardt2022a,Steinhardt2022b,Rusakov2023}.  These inferred gas temperatures are also consistent with dust temperature estimates lying above Galactic temperatures, both at intermediate redshift \citep{Magnelli2014} and at high redshift \citep{Sommovigo2022}.  In effect, nearly every star-forming galaxy is found to have a bottom-lighter IMF than the Milky Way.  It is only once a galaxy quenches that the IMF temperature stars to drop to Galactic temperatures or even below \citep{Steinhardt2023}.  Thus, if nearly the entire stellar population formed with a bottom-light IMF, measurements based on the stellar population in quiescent galaxies should produce the same result: a bottom-light IMF.  

However, instead the opposite appears to be true.  Spectral lines believed to be associated with T dwarfs indicate that the IMF corresponding to the bulk of the present-day stellar population in elliptical galaxies is instead bottom-heavy \citep{Conroy2012}.  The same appears to be true for the Gaia-Enceladus population \citep{Hallakoun2021}, which is believed to have formed at relatively high redshift.  This would not be possible if the bulk of star formation occurs with a bottom-light IMF, as implied by both observation and theory for higher-redshift star-forming galaxies.  That is, although an aging stellar population loses its most massive stars, there is no aging process which can convert a stellar population formed with a bottom-light IMF into one that appears bottom-heavy.  Similarly, there is no merger process which can combine bottom-light stellar populations (or even populations produced with a Galactic IMF) into a bottom-heavy one.  The only way to change a bottom-light population to a bottom-heavy one is to add a significant number of additional stars with a bottom-heavy IMF, enough to dominate the resulting merged population.  In \S~\ref{sec:rsfg}, a mechanism is instead proposed for star formation to continue with a bottom-heavy IMF past the point of apparent quenching.

\section{Red Star-Forming Galaxies}
\label{sec:rsfg}

The puzzles described in the previous section might have a common solution under the following scenario: perhaps galaxies which appear to have quenched are instead continuing to form stars efficiently, but with a bottom-heavy IMF that does not include any O or B stars, so that the galaxy will appear red rather than blue.  A typical galaxy would form the last factor of $\sim 10$ of its stellar mass in this manner, which we here term a {\em red star-forming galaxy} (RSFG).  Here, a scenario in which such an RSFG phase might not only be possible, but typical, is proposed.

The IMF is set by a combination of two different processes required for star formation.  First, molecular clouds must collapse, a process often characterized by the Jeans instability and associated Jeans mass \citep{Jeans1902},
\begin{equation}
M_J = \frac{\pi c_s^3}{6G^{3/2}\rho^{1/2}}, 
\end{equation} 
where $c_s$ is the speed of sound within a cloud of density $\rho$.  Clouds above the Jeans mass will collapse and form stars.  For high-redshift star-forming galaxies, the Jeans mass is typically $10^3-10^4 M_\odot$.  This does not produce a single, supermassive star, but rather as the cloud condenses, it will fragment into smaller clumps which produce individual stars.  This fragmentation process, also dependent upon $c_s$ \citep{LyndenBell1976,Larson1985}, is what sets the IMF.

However, there is a second physical limit: it is impossible to produce a star more massive than the collapsing cloud, so stars above the Jeans mass will not be formed.  Thus, if the Jeans mass drops below the masses of, e.g., O stars, then no more O stars will be formed regardless of the fragmentation scale.  In practice, this is almost certainly not a sharp cutoff.  The Jeans mass will vary for different molecular clouds within a galaxy due to variations in temperature, density, and metallicity \citep{Jeans1902,Hopkins2012,Chabrier2014}.  Further, the Jeans mass itself is an approximation to a characteristic scale for collapse rather than a rigorously-derived cutoff.  Nevertheless, it is reasonable to describe the size of molecular clouds as having some high-mass cutoff, and this cutoff is likely reasonably approximated by $M_J$.

For high-redshift star-forming galaxies, $M_J$ is typically $10^3 - 10^4 M_\odot$, so that the IMF is primarily determined by fragmentation.  However, for a Galactic molecular cloud, the speed of sound is $\sim 0.2 km/s$ \citep{Krumholz2008}, and at a density of $\sim 10^2 cm^{-3}$ \citep{Heyer2015}, the Jeans mass is only $M_J \sim 10 M_\odot$\footnote{Because of variability in conditions within Galactic molecular clouds, this does not mean that $M_J \sim 10 M_\odot$ is a hard cutoff for the maximum mass of newly-formed stars.  However, it is unlikely that this variability allows the formation of clouds far above the average $M_J$.}. This proximity to the mass of massive stars was one of the reasons astronomers first concluded that molecular clouds were the site of star formation \citep{Bergin2007}.  As the temperature continues to drop and metallicity increases, the speed of sound and thus the Jeans mass will drop further.  Thus, at some point, it will no longer be possible to form the most massive O stars.  As $M_J$ continues to drop, it will then be impossible to form any O stars, then any O or B stars, etc.  Within $\sim 500$ Myr, just as with a truly quenched stellar population, such a galaxy would change color from blue to red, due to the absence of massive stars in its stellar population.  However, during this RSFG phase, the galaxy will continue to form stars efficiently.
  
In a typical star-forming galaxy, the young stellar population is believed to provide the dominant contribution to molecular cloud temperature.  A wide variety of mechanisms have been proposed, including stellar feedback from massive stars \citep{Lopez2014}, supernova heating \citep{Kobayashi2009}, cosmic rays \citep{Papadopoulos2010,Jermyn2018} and subsequent AGN growth and feedback \citep{Fabian2012}.  However, all of these processes are driven by either the most massive main-sequence stars or their deaths, so in an RSFG, none of them can happen.  Instead, there is no barrier to a rapid increase in star-formation rate, allowing a RSFG to rapidly use up all of its available gas by forming low-mass stars.  Thus, a RSFG will simply continue to form stars until its exhausts all of the available gas and dust, at which point it will truly become quenched for lack of additional material out of which the galaxy can form more stars.  The possibility of such a feedback-free phase allowing rapid stellar mass growth has also been proposed at high redshifts in an attempt to explain the excess of early, luminous galaxies \citep{Dekel2023}.  If a galaxy has a high gas mass to stellar mass ratio $M_g / M_*$, this process could potentially form the bulk of the final stellar population by the time galaxies truly quench.  This would be consistent with the observed decrease in $M_g / M_*$ with increasing $M_*$ observed in the local Universe \citep{Parkash2018}.  

\subsection{Post-Starburst Galaxies and Multiple Quenching Modes}
\label{subsec:poststarburst}

There is strong observational evidence for galaxies with spectra dominated by A stars ranging from the local Universe to high redshift \citep{Zabludoff1996,Dressler1999,French2015,Alatalo2016}.  Many of these are also best fit with a relative young stellar population \citep{Quintero2004,Wild2009}, or with a composite population comprised of a younger (A-star-dominated) and older (K-star) population, leading to an alternative description as `K+A' galaxies \citep{French2021}.  

The standard interpretation is that these galaxies are in a rapid transition between having a high star-formation rate and quiescence, leading to their description as post-starburst galaxies \citep{Spinrad1973,Dressler1983,Couch1987}.  Spectra are also consistent with this picture, exhibiting strong H$\delta$ emission indicating recent star formation \citep{Worthey1997} but a lack of [O{\sc ii}]$\lambda$3727 or H$\alpha$ which would indicate current star formation.  Many K+A galaxies are also known to be associated with mergers \citep{French2021}, suggesting that they are environmentally driven rather than a phase in the secular evolution of a typical galaxy.

More precisely, though, [O{\sc ii}]$\lambda$3727 is an indicator of O and B stars, which in turn due to their short lifetimes are considered an indicator of current star formation.  Similarly, H$\delta$ emission indicates the presence of A stars, and a strong Balmer break indicates that light from A stars dominates the stellar population.  Finally, additional emission in the infrared indicates a large population of lower-mass stars.  

\subsection{RSFGs as Post-Starburst Galaxies}

However, the same set of observational signatures could be produced by RSFGs.  The light from a galaxy is dominated by its existing stellar population, and therefore the key spectral properties of an RSFG come from the principal features of its stellar population:
\begin{itemize}
    \item{An absence of O and B stars, due to an inability to form massive stars.}
    \item{A stellar population dominated by A stars, including ones which are actively being formed.}
    \item{A bottom-heavy initial mass function, and thus a stellar population with a high fraction of low-mass stars.}
\end{itemize}
\begin{figure}
    \centering
    \includegraphics[width=0.45\textwidth]{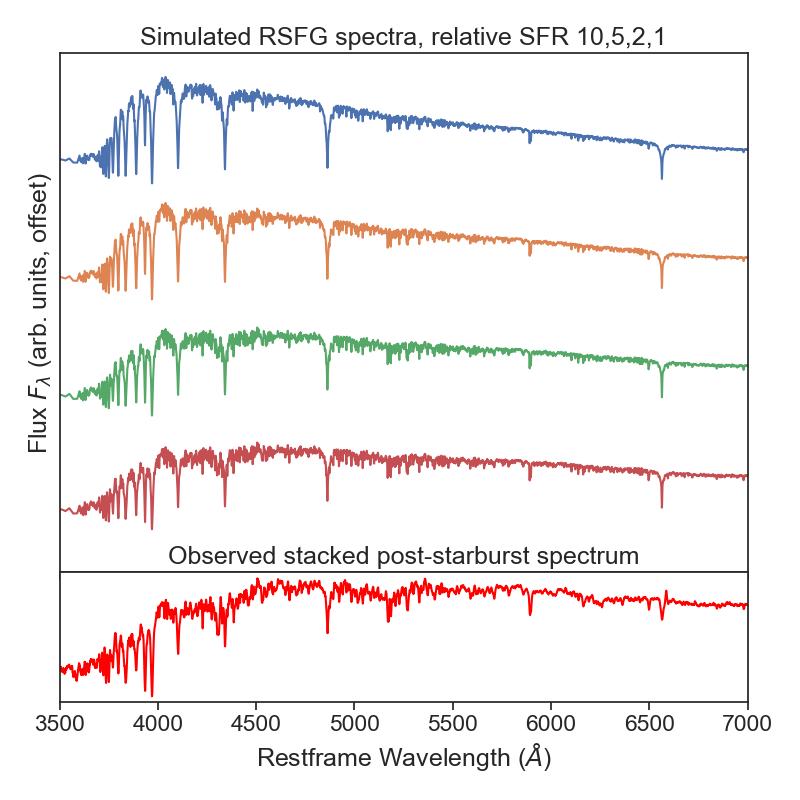}
    \label{fig:rsfgspec}
    \caption{Four sample possible spectra of red star-forming galaxies generated using FSPS \citep{Conroy2009b}.  Each combines a 1Gyr old stellar population formed as a blue star-forming galaxy with a 100 Myr old red star-forming galaxy stellar population with a $3 M_\odot$ mass cutoff and several choices of star-formation rate.  The red star-forming galaxy SFR increases from bottom to top (relative SFR of 1, 2, 5, and 10), resulting in a spectral energy distribution increasingly dominated by A stars.  The spectrum has many features in common with those of post-starburst (or K+A) galaxies, as shown in a stacked spectrum at bottom (Nielsen et al., in prep.).}
\end{figure}

The absence of O and B stars lead to a red galaxy without [O{\sc ii}]$\lambda$3727 nebular emission lines (cf. \citet{French2021}).  Because the light is dominated by a recently-formed A star population, there will be a strong Balmer break and H$\delta$ emission.  Finally, because the IMF is bottom-heavy, indicators sensitive to low-mass stars will indicate a large low-mass stellar population.  These are also the key selection criteria for post-starburst galaxies, suggesting the possibility that red star-forming galaxies might be hidden within post-starburst selections.  

In an attempt to explore this possibility, synthetic RSFG spectra can be compared with observed post-starburst galaxies.  Detailed simulations will be required to determine key properties such as the length of an RSFG phase, and feedback mechanisms that limit star formation rates.  For illustrative purposes, one might consider a galaxy which has been unable to form O stars for the past $10^9$ yr.  If the star-formation history is otherwise typical of a galaxy on the blue star-forming main sequence, then the luminosity-weighted age of the stellar population still in production should be $\sim 10^8$ yr.

Using Flexible Stellar Population Synthesis (FSPS; \citet{Conroy2009b}), a synthetic spectrum of an RSFG was constructed by combining a 1 Gyr old stellar population formed in a typical blue star-forming phase with a 100 Myr old stellar population with a $3 M_\odot$ mass cutoff.  Several possible choices of SFR for the younger stellar population are considered over an order of magnitude, as examples of possible spectra that could be produced depending upon the efficiency of a red star-forming phase (Fig. \ref{fig:rsfgspec}, top).  This SED is comparable to those of post-starburst galaxies (\citealt{French2021}; Fig. \ref{fig:rsfgspec}, bottom) and also has similar UV features to spectra of high-redshift quiescent galaxy candidates from JWST \citep{Heintz2023}.  

Although it would also be tempting to compare detailed emission and absorption lines with post-starburst galaxies, nebular emission in FSPS is only based on a Galactic IMF and related physical assumptions. At a minimum, then, it would be necessary to recalculate these lines using a spectral synthesis code such as \texttt{Cloudy} \citep{Chatzikos2023}.  However, the answer will still depend upon the choice of other parameters such as HI and HII gas mass and other feedback mechanisms.  

A common approach to this problem in photometric template fitting is to attempt to span all possible combinations, hoping that the correct physical model will prove to be the best fit and that physically implausible combinations will never succeed.  Unfortunately, this is rarely entirely successful, as evidenced by the numerous galaxies found in photometric property catalogs with stellar populations that appear to be older than the age of the Universe.  So, perhaps the best solution is to first use numerical galaxy simulations to determine what combinations of parameters can be sustainable and which are disallowed by feedback or other astrophysics, then to run those parameters through spectral synthesis codes to produce lines corresponding to the correct radiative transfer astrophysics.  The synthetic spectra shown here likely have several key features in common with the eventual result due to having a similar stellar population, but might not correspond to physically plausible RSFG spectra, and certainly do not span the full range of possible RSFG spectra.

Broadly, though, these synthetic spectra suggest that a RSFG would strongly resemble a post-starburst galaxy and RSFGs could comprise an interloper population within post-starburst samples.  If so, the interpretation of this subset of K+A or E+A galaxies would change.  Rather than being a composite of an older stellar population with one that formed rapidly and recently quenched, these would instead be red star-forming galaxies, a phase which follows blue star formation in the evolution of a typical galaxy.  

If it were possible to separate the two populations, the frequency of K+A galaxies could then be used to determine the approximate duration of RFSG growth.  An additional complication is that although an RSFG will not efficiently form dust, it will similarly be unable to efficiently destroy dust.  In a blue star-forming galaxy, the dominant processes which destroy dust include supernova shocks, radiation from massive stars, and AGN feedback.  All of these would be strongly suppressed or entirely non-existent in an RSFG, so residual dust from the end of the blue star-forming phase would potentially continue to obscure RSFGs for an extended period.  

Thus, it is possible that rather than galaxies evolving from a blue star-forming galaxy directly to a K+A state, there would be an intermediate dusty RSFG phase with strong extinction and in which the emission lines are substantially obscured.  Depending upon the rate at which dust is destroyed and whether the dust is thicker in regions of ongoing low-mass star formation, such galaxies might be selected either as star-forming or quiescent in a color-color selection such as UVJ or NUVrJ.  However, an RSFG such as in Fig. \ref{fig:rsfgspec} will always be selected as quiescent from a color-color diagram.

\section{Red Star-Forming Galaxies as Solution to Existing Puzzles}
\label{sec:solution}

In \S~\ref{sec:puzzles}, three observational results were described which are difficult to explain within the standard scenario for galaxy evolution:
\begin{itemize}
    \item{An apparent $M_*/M_{BH} \sim 30$ for quenching galaxies, compared with a bulge $M_*/M_{BH} \sim 300$ in the local Universe.}
    \item{A $\sim 1$ dex gap between the stellar masses of local, massive galaxies and the most massive star-forming galaxies.}
    \item{A bottom-heavy IMF in early-type galaxies compared with a bottom-light IMF in star-forming galaxies.}
\end{itemize}
Here, it is shown that all three sets of observations can be reconciled within the RSFG scenario.

\subsection{RSFGs and Stellar Mass Growth}

An RSFG will continue to produce stars until either it runs out of gas or the Jeans mass drops below the minimum possible stellar mass of around $0.08 M_\odot$.  Because massive stars with short lifetimes cannot be produced, their deaths will not trigger additional supermassive black hole growth from supernovae.  Thus, this provides an increase in stellar mass without an accompanying increase in black hole mass.  If the stellar mass grows by a factor of 10 during this period, then $M_* / M_{BH} = 30$ at the conclusion of the blue star-forming phase will yield $M_* / M_{BH} = 300$ at the end of the RSFG phase, consistent with local observations \citep{Magorrian1998,Haring2004,McConnell2013}.  Such growth is plausible for a galaxy with high gas mass to stellar mass ratio during its blue star-forming phase.

The most massive blue star-forming galaxies in the COSMOS2020 catalog \citep{Weaver2022,Rusakov2023} have $M_* \sim 10^{11.5} M_\odot$.  So, the same factor of 10 growth in stellar mass for these galaxies would yield $M_* = 10^{12.5} M_\odot$, comparable to the most massive local galaxies such as M87 \citep{Gebhardt2009} and IC 1101 \citep{Dullo2017}.  Moreover, it would do so without an increase in supermassive black hole mass, so that the most massive quasars in the \citet{Shen2011} SDSS DR11 catalog, with $M_{BH} \sim 10^{9.8} M_\odot$, would have those same masses today.  This, too, is consistent with M87 and IC 1101, but would not be consistent with their having grown that additional stellar mass via mergers, since those would also produce accompanying supermassive black hole mass growth.

\subsection{RSFGs and IMF Evolution}

Blue star-forming galaxies are both predicted and observed to have bottom-light initial mass functions.  Thus, if these populations merely age or merge, the resulting stellar population must continue to be bottom-light.  However, a RSFG forms stars at lower temperatures than Galactic molecular clouds.  Simulations find that cosmic ray heating alone would only produce Galactic temperatures of $\sim 5$ K \citep{Papadopoulos2010}, and observations find temperatures in the $\sim 10$ K range \citep{Greenberg1996}.

Further, because there is no feedback from massive stars or their deaths, temperatures should continue to decline as star formation proceeds.  Thus, since lower temperatures and higher metallicities produce bottom-heavier IMFs \citep{Larson1985,Kroupa2002,Jermyn2018} an RSFG will form stars with a bottom-heavy IMF.

\citet{Conroy2012} find that in local red elliptical galaxies, the mass-to-light ratio increases with increasing metallicity.  At [Z/H] near 0.2, the K-band mass-to-light ratio is between 1.6 times and twice the Galactic ratio.  Using the IMF given by Eq. \ref{eq:imf}, similar mass-to-light ratios would be produced by gas temperatures of $T \sim 5-9$K.  This combination of lower temperatures and higher metallicities is, at least qualitatively, what one would expect for the long-term future of a galaxy like our own.

The Galactic Jeans mass is $M_J \sim 10 M_\odot$ at $T \sim 20 K$.  So, at $T \sim 5-9$K, keeping all other variables fixed, the Jeans mass would drop to $M_J \sim 0.8 - 4.6 M_\odot$.  If this is the maximum stellar mass at formation, then it is plausible that such a galaxy would be unable to form B stars.

If the stellar mass grows by a factor of 10 during the RSFG phase, the resulting stellar population will be dominated by stars produced with these bottom-heavy IMFs.  Thus, once the galaxy has truly quenched, it will be consistent with observations of early-type galaxies with bottom-heavy IMFs \citep{Conroy2012}.  Similarly, if the RSFG phase is relatively rapid due to a lack of feedback that would otherwise limit star formation, it should be possible to produce these bottom-heavy stellar populations shortly after the first galaxies turn from blue to red.  

Galactic molecular clouds have a mass distribution $n(M) \propto M^{-1.60\pm 0.06}$ \citep{Gomez2014}, which is believed to have negligible temperature dependence \citep{Elmegreen2011}.  Assuming that this power law continues to hold for clouds at $M < 10 M_\odot$, where current observations have not been able to measure the distribution, the SFR should scale approximately as $\int_{M_J}^{\infty}n(M)dM \propto M_J^{-0.60 \pm 0.06} \propto T^{-0.90 \pm 0.09}$.  For a high-redshift galaxy turning off from an sSFR of $10^-8.5 /yr$ at $T = 30$K and $z = 4$, it would take under 1 Gyr to grow the stellar mass by a factor of 10 in a RSFG phase.  

Thus, this scenario is also consistent with a bottom-heavy Gaia-Enceladus population \citep{Hallakoun2021} which appears to be 8-11 Myr old \citep{Myeong2018}. However, a galaxy which is only quenching at $z = 0$ would take $\sim 1$ dex longer to grow by a similar factor.  In both cases, the RSFG phase would be shorter than the blue star-forming phase.

\section{Discussion}
\label{sec:discussion}

Galaxies are observed to have a broadly bimodal color distribution, dividing them into red and blue populations.  In the standard picture of galaxy formation and evolution, blue galaxies are star-forming, while most red galaxies (apart from a dusty star-forming population) are thought to have quenched.  However, a variety of observations challenge this scenario by indicating that following the blue star-forming phase, galaxies must significantly increase in stellar mass and change from a stellar population formed with an IMF bottom-lighter than the Milky Way to one that is bottom-heavier.  Here, a new model is proposed in which these problems are solved via a red star-forming galaxy (RSFG) phase, in which most of final stellar mass of a galaxy is formed.  These galaxies would closely resemble post-starburst galaxies, and would likely be an interloper population in current post-starburst samples.

\subsection{Previous Analysis and Models}
The possibility that the $M_*/M_{BH}$ ratio might need to increase by a factor of 10 between blue star-forming galaxies and local red galaxies was previously reported \citep{Steinhardt2014b}.  The same study proposed a model in which during the last stages of main sequence star formation, the duty cycle of the central supermassive black hole would be 100\%.  Thus, quasar accretion would `hide' the last $\sim 90\%$ of main sequence star formation, since the light would be dominated by the AGN.  

A similar mismatch was rediscovered more recently at $z \sim 1-2$ \citep{Farrah2023a}.  At these redshifts, $(1+z)^3$ is similar to the required difference in mass ratio.  \citet{Farrah2023b} proposed that supermassive black holes could contain a form of dark energy, so that the black hole energy density would scale differently than the stellar mass energy density as the Universe expands.  However, this would likely produce a redshift-dependent $M_*/M_{BH}$ ratio rather than the approximately constant $M_*/M_{BH}$ in Fig. \ref{fig:downsizing}.

A far more mundane explanation would be errors in mass estimation.  A systematic overestimate of black hole virial masses or underestimate of stellar masses would also provide a solution.  Both are historically difficult to measure and rely on the assumption that scaling relations calibrated in the local Universe apply even at high redshift.  

The new constraints come from measurements of the initial mass function.  The mismatch between the bottom-light IMFs both predicted and observed for high-redshift star-forming galaxies and the bottom-heavy IMFs in local red ellipticals cannot be explained by previous models.  Star formation hidden by quasar accretion would take place in an environment heated by the AGN, and thus would be bottom-light.  A black hole mass density which evolves like that of dark energy would similarly be unable to change the stellar mass distribution.

These IMF measurements also place additional constraints on $M_*/M_{BH}$ ratios.  A bottom-light IMF produces a lower mass-to-light ratio than the Galactic IMF and thus results in a lower $M_*/M_{BH}$ at turnoff (Fig. 1, purple).  Further, not only is a bottom-light IMF expected at high redshift (cf. \citealt{LyndenBell1976,Larson1985,Jermyn2018}, but the reduced stellar masses provide a possible solution to the puzzle of `impossibly' massive galaxies \citep{Steinhardt2016,Labbe2023,BoylanKolchin2023,Steinhardt2023}.  Thus, although stellar masses continue to have $\sim 0.5$ dex uncertainties, it is unlikely that they have been significantly \emph{underestimated} for the most massive star-forming galaxies, which are already too massive under $\Lambda$CDM.  

Virial black hole masses also have large uncertainties, and the scaling relations based on lower-redshift reverberation mapping \citep{McLure2004,Vestergaard2006} have large scatter for higher-redshift quasars \citep{Shen2024}.  The luminosities of these quasars provide a strong astrophysical lower bound on their black hole masses, since they cannot exceed their Eddington luminosities.  Combined with the expectation that due to their IMF, stellar masses are likely overestimated, this bounds $M_*/M_{BH}$ below the local red elliptical value, even accounting for uncertainties.  Thus, each of the previous proposed explanations is inconsistent with current observations.  A new model, such as the possibility of red star-forming galaxies, is thus required.

\subsection{Falsifiable Predictions of Red Star-Forming Galaxies}

The value of a new model lies in a combination of its explanatory and predictive capabilities.  As with many astrophysical ideas, the RSFG model is driven by several puzzling observations for which there had previously been no satisfactory explanation.  Thus, there is significant explanatory power in a relatively simple model which can simultaneously explain all of the differences between high-redshift blue galaxies and local red galaxies.

There are two significant difficulties in finding additional falsifiable predictions.  First, the observed light is always dominated by rare, massive stars.  
Current SFR indicators are predominantly instead measures of O star flux, scaled based on an assumed IMF and the lifetime of O stars to an average SFR for the galaxy.  If the IMF does not produce O stars, these indicators will fail, and new indicators must be developed which would be sensitive to proposed RSFG star formation.  A model which is distinguished primarily by a distinct low-mass stellar population is therefore very difficult to properly test, even more so because seeking an explanation for the limited existing observations of low-mass stellar populations was part of the rationale for its development.   

Second. there is a degeneracy between the IMF and star formation history (SFH), since different combinations of SFH and IMF can produce an identical current stellar population.  In particular, the same stellar population with an apparentl surplus of low-mass stars and A stars and without O stars, could be produced both with a short, very high SFR burst followed by rapid quenching and enough elapsed time to lose the most massive stars and by an RSFG phase.  This degeneracy is also the main reason that although a non-Universal IMF has long been predicted, it has been so difficult to find strong observational evidence.  However, there are additional differences between RSFGs and post-starburst galaxies, leading to predictions which could be potentially tested in the near future.

\subsection{Resolved Spectroscopy of Post-Starburst Galaxies}

A potential test comes from comparing stellar populations in galactic centers and disks.  It has long been known that the Galactic center contains a lower-metallicity, older stellar population \citep{Baade1944}.  Estimates of the IMF and gas temperatures of high-redshift galaxies are consistent with the hypothesis that the cores of all typical galaxies form earlier, and with a bottom-lighter IMF, than disks \citep{Steinhardt2023}.  Bulge-disk decompositions similarly are consistent with the possibility that the core forms earlier, and that the star-forming main sequence is only a disk-forming main sequence \citep{Abramson2014}.  Recently, such a decomposition has become possible at higher redshift as well \citep{GimenezArteaga2023}, so that it might be possible to resolve the cores of star-forming galaxies independently from their disks.

If this earlier core formation scenario is correct, due to higher densities in central regions the Jeans mass should fall earlier than in the disk.  At some point, then, the core a galaxy might resemble an RSFG while the disk is still blue and star-forming.  This is the opposite of a merger-driven starburst scenario, in which star formation instead should quench outside-in \citep{Rowlands2018}.  In resolved spectroscopy, the core of an idealized RSFG at this stage should resemble a K+A galaxy and the disk should have a typical blue star-forming spectrum.  Finding such galaxies would be strong evidence for the evolutionary history proposed here. 
 Because molecular clouds within a galaxy can have a range of conditions, with sufficient resolution a more complex picture might emerge instead.

An initial test can be conducted using resolved spectroscopy from the Mapping Nearby Galaxies at Apache Point Observatory (MaNGA; \citealt{Bundy2015}) survey.  Of 49 MaNGA post-starburst galaxies identified as being post-merger, 22 (44.9\%) showed outside-in quenching while only 3 (6.1\%) were quenching inside-out and 6 (12.2\%) exhibited more complex behavior with no clear trend \citep{Rowlands2018,Li2023}.  However, in a control sample of 156 post-starburst galaxies without clear post-merger signatures, more quenched inside-out (40; 25.6\%) than outside-in (30; 19.6\%), with 52 (33.3\%) exhibiting no clear trend \citep{Li2023}.  

This difference is consistent with the hypothesis that K+A galaxies can be produced via two distinct mechanisms, one from mergers and another from secular evolution.  Merger-driven galaxies would properly be described as post-starburst, while secular evolution would instead produce an RSFG.  Additional resolved spectroscopy, ideally for post-starburst galaxies at higher redshifts and which show no indications of recent merger activity, would allow a stronger test of whether intrinsically red galaxies truly form more stellar mass than intrinsically blue ones.   

An RSFG mechanism for producing post-starburst spectra might also provide an explanation for the observed overrepresentation of tidal disruption events (TDEs) in post-starburst galaxies by a factor of$\sim 25-50$ \citep{LawSmith2017,Hammerstein2021}.  Because these galaxies would be actively forming low-mass stars, particularly in the central region of the galaxies where the Jeans mass is lowest, they should exhibit a significantly higher TDE rate than in a merger-driven scenario.  Thus, an additional prediction is that the TDE overrepresentation should be concentrated in the post-starburst galaxies which do not show clear evidence of being merger-driven, as those are the strongest candidates for RSFGs.

Finally, if a RSFG phase is truly the successor to blue star formation, this should be evident from mass measurements.  One of the effects of downsizing is that at any redshift, the most massive blue star-forming galaxies will no longer be blue star-forming galaxies at lower redshift.  Thus, if RSFGs are hidden in post-starburst selections, a substantial population of post-starburst galaxies must lie in a well-defined mass range, spanning the most massive blue star-forming galaxies and the least massive quiescent ones.  If post-starburst galaxies are instead caused by merger-driven rejuvenation of quiescent galaxies, their mass distribution should be more similar to the full quiescent mass distribution.

The author would like to thank Thorbj\o rn Clausen, Avishai Dekel, Decker French, Aigen Li, Gustav Lindstad, Dani Maoz, Conor McPartland, Bahram Mobasher, Emma Nielsen, Vadim Rusakov, Albert Sneppen, Scott Tremaine, Darach Watson, and John Weaver for helpful comments.

\bibliographystyle{aasjournal}
\bibliography{refs.bib} 

\begin{thebibliography}{}
\expandafter\ifx\csname natexlab\endcsname\relax\def\natexlab#1{#1}\fi
\providecommand{\url}[1]{\href{#1}{#1}}
\providecommand{\dodoi}[1]{doi:~\href{http://doi.org/#1}{\nolinkurl{#1}}}
\providecommand{\doeprint}[1]{\href{http://ascl.net/#1}{\nolinkurl{http://ascl.net/#1}}}
\providecommand{\doarXiv}[1]{\href{https://arxiv.org/abs/#1}{\nolinkurl{https://arxiv.org/abs/#1}}}

\bibitem[{{Abramson} {et~al.}(2014){Abramson}, {Kelson}, {Dressler}, {Poggianti}, {Gladders}, {Oemler}, \& {Vulcani}}]{Abramson2014}
{Abramson}, L.~E., {Kelson}, D.~D., {Dressler}, A., {et~al.} 2014, \apjl, 785, L36, \dodoi{10.1088/2041-8205/785/2/L36}

\bibitem[{{Alatalo} {et~al.}(2016){Alatalo}, {Cales}, {Rich}, {Appleton}, {Kewley}, {Lacy}, {Lanz}, {Medling}, \& {Nyland}}]{Alatalo2016}
{Alatalo}, K., {Cales}, S.~L., {Rich}, J.~A., {et~al.} 2016, \apjs, 224, 38, \dodoi{10.3847/0067-0049/224/2/38}

\bibitem[{{Antwi-Danso} {et~al.}(2023){Antwi-Danso}, {Papovich}, {Leja}, {Marchesini}, {Marsan}, {Martis}, {Labb{\'e}}, {Muzzin}, {Glazebrook}, {Straatman}, \& {Tran}}]{AntwiDanso2023}
{Antwi-Danso}, J., {Papovich}, C., {Leja}, J., {et~al.} 2023, \apj, 943, 166, \dodoi{10.3847/1538-4357/aca294}

\bibitem[{{Baade}(1944)}]{Baade1944}
{Baade}, W. 1944, \apj, 100, 137, \dodoi{10.1086/144650}

\bibitem[{{Bate}(2005)}]{Bate2005}
{Bate}, M.~R. 2005, \mnras, 363, 363, \dodoi{10.1111/j.1365-2966.2005.09476.x}

\bibitem[{{Bergin} \& {Tafalla}(2007)}]{Bergin2007}
{Bergin}, E.~A., \& {Tafalla}, M. 2007, \araa, 45, 339, \dodoi{10.1146/annurev.astro.45.071206.100404}

\bibitem[{{Boylan-Kolchin}(2023)}]{BoylanKolchin2023}
{Boylan-Kolchin}, M. 2023, Nature Astronomy, \dodoi{10.1038/s41550-023-01937-7}

\bibitem[{{Bundy} {et~al.}(2015){Bundy}, {Bershady}, {Law}, {Yan}, {Drory}, {MacDonald}, {Wake}, {Cherinka}, {S{\'a}nchez-Gallego}, {Weijmans}, {Thomas}, {Tremonti}, {Masters}, {Coccato}, {Diamond-Stanic}, {Arag{\'o}n-Salamanca}, {Avila-Reese}, {Badenes}, {Falc{\'o}n-Barroso}, {Belfiore}, {Bizyaev}, {Blanc}, {Bland-Hawthorn}, {Blanton}, {Brownstein}, {Byler}, {Cappellari}, {Conroy}, {Dutton}, {Emsellem}, {Etherington}, {Frinchaboy}, {Fu}, {Gunn}, {Harding}, {Johnston}, {Kauffmann}, {Kinemuchi}, {Klaene}, {Knapen}, {Leauthaud}, {Li}, {Lin}, {Maiolino}, {Malanushenko}, {Malanushenko}, {Mao}, {Maraston}, {McDermid}, {Merrifield}, {Nichol}, {Oravetz}, {Pan}, {Parejko}, {Sanchez}, {Schlegel}, {Simmons}, {Steele}, {Steinmetz}, {Thanjavur}, {Thompson}, {Tinker}, {van den Bosch}, {Westfall}, {Wilkinson}, {Wright}, {Xiao}, \& {Zhang}}]{Bundy2015}
{Bundy}, K., {Bershady}, M.~A., {Law}, D.~R., {et~al.} 2015, \apj, 798, 7, \dodoi{10.1088/0004-637X/798/1/7}

\bibitem[{{Chabrier}(2003)}]{Chabrier2003}
{Chabrier}, G. 2003, PASP, 115, 763, \dodoi{10.1086/376392}

\bibitem[{{Chabrier} {et~al.}(2014){Chabrier}, {Hennebelle}, \& {Charlot}}]{Chabrier2014}
{Chabrier}, G., {Hennebelle}, P., \& {Charlot}, S. 2014, \apj, 796, 75, \dodoi{10.1088/0004-637X/796/2/75}

\bibitem[{{Conroy} \& {van Dokkum}(2012)}]{Conroy2012}
{Conroy}, C., \& {van Dokkum}, P.~G. 2012, \apj, 760, 71, \dodoi{10.1088/0004-637X/760/1/71}

\bibitem[{{Conroy} \& {Wechsler}(2009)}]{Conroy2009b}
{Conroy}, C., \& {Wechsler}, R.~H. 2009, \apj, 696, 620, \dodoi{10.1088/0004-637X/696/1/620}

\bibitem[{{Couch} \& {Sharples}(1987)}]{Couch1987}
{Couch}, W.~J., \& {Sharples}, R.~M. 1987, \mnras, 229, 423, \dodoi{10.1093/mnras/229.3.423}

\bibitem[{{Dav{\'e}}(2008)}]{Dave2008}
{Dav{\'e}}, R. 2008, \mnras, 385, 147, \dodoi{10.1111/j.1365-2966.2008.12866.x}

\bibitem[{{Dekel} {et~al.}(2023){Dekel}, {Sarkar}, {Birnboim}, {Mandelker}, \& {Li}}]{Dekel2023}
{Dekel}, A., {Sarkar}, K.~C., {Birnboim}, Y., {Mandelker}, N., \& {Li}, Z. 2023, \mnras, 523, 3201, \dodoi{10.1093/mnras/stad1557}

\bibitem[{{Dressler} \& {Gunn}(1983)}]{Dressler1983}
{Dressler}, A., \& {Gunn}, J.~E. 1983, \apj, 270, 7, \dodoi{10.1086/161093}

\bibitem[{{Dressler} {et~al.}(1999){Dressler}, {Smail}, {Poggianti}, {Butcher}, {Couch}, {Ellis}, \& {Oemler}}]{Dressler1999}
{Dressler}, A., {Smail}, I., {Poggianti}, B.~M., {et~al.} 1999, \apjs, 122, 51, \dodoi{10.1086/313213}

\bibitem[{{Dullo} {et~al.}(2017){Dullo}, {Graham}, \& {Knapen}}]{Dullo2017}
{Dullo}, B.~T., {Graham}, A.~W., \& {Knapen}, J.~H. 2017, \mnras, 471, 2321, \dodoi{10.1093/mnras/stx1635}

\bibitem[{{Elmegreen}(2011)}]{Elmegreen2011}
{Elmegreen}, B.~G. 2011, \apj, 737, 10, \dodoi{10.1088/0004-637X/737/1/10}

\bibitem[{{Event Horizon Telescope Collaboration} {et~al.}(2019){Event Horizon Telescope Collaboration}, {Akiyama}, {Alberdi}, {Alef}, {Asada}, {Azulay}, {Baczko}, {Ball}, {Balokovi{\'c}}, {Barrett}, {Bintley}, {Blackburn}, {Boland}, {Bouman}, {Bower}, {Bremer}, {Brinkerink}, {Brissenden}, {Britzen}, {Broderick}, {Broguiere}, {Bronzwaer}, {Byun}, {Carlstrom}, {Chael}, {Chan}, {Chatterjee}, {Chatterjee}, {Chen}, {Chen}, {Cho}, {Christian}, {Conway}, {Cordes}, {Crew}, {Cui}, {Davelaar}, {De Laurentis}, {Deane}, {Dempsey}, {Desvignes}, {Dexter}, {Doeleman}, {Eatough}, {Falcke}, {Fish}, {Fomalont}, {Fraga-Encinas}, {Freeman}, {Friberg}, {Fromm}, {G{\'o}mez}, {Galison}, {Gammie}, {Garc{\'\i}a}, {Gentaz}, {Georgiev}, {Goddi}, {Gold}, {Gu}, {Gurwell}, {Hada}, {Hecht}, {Hesper}, {Ho}, {Ho}, {Honma}, {Huang}, {Huang}, {Hughes}, {Ikeda}, {Inoue}, {Issaoun}, {James}, {Jannuzi}, {Janssen}, {Jeter}, {Jiang}, {Johnson}, {Jorstad}, {Jung}, {Karami}, {Karuppusamy}, {Kawashima}, {Keating}, {Kettenis}, {Kim}, {Kim}, {Kim},
  {Kino}, {Koay}, {Koch}, {Koyama}, {Kramer}, {Kramer}, {Krichbaum}, {Kuo}, {Lauer}, {Lee}, {Li}, {Li}, {Lindqvist}, {Liu}, {Liuzzo}, {Lo}, {Lobanov}, {Loinard}, {Lonsdale}, {Lu}, {MacDonald}, {Mao}, {Markoff}, {Marrone}, {Marscher}, {Mart{\'\i}-Vidal}, {Matsushita}, {Matthews}, {Medeiros}, {Menten}, {Mizuno}, {Mizuno}, {Moran}, {Moriyama}, {Moscibrodzka}, {M{\"u}ller}, {Nagai}, {Nagar}, {Nakamura}, {Narayan}, {Narayanan}, {Natarajan}, {Neri}, {Ni}, {Noutsos}, {Okino}, {Olivares}, {Ortiz-Le{\'o}n}, {Oyama}, {{\"O}zel}, {Palumbo}, {Patel}, {Pen}, {Pesce}, {Pi{\'e}tu}, {Plambeck}, {PopStefanija}, {Porth}, {Prather}, {Preciado-L{\'o}pez}, {Psaltis}, {Pu}, {Ramakrishnan}, {Rao}, {Rawlings}, {Raymond}, {Rezzolla}, {Ripperda}, {Roelofs}, {Rogers}, {Ros}, {Rose}, {Roshanineshat}, {Rottmann}, {Roy}, {Ruszczyk}, {Ryan}, {Rygl}, {S{\'a}nchez}, {S{\'a}nchez-Arguelles}, {Sasada}, {Savolainen}, {Schloerb}, {Schuster}, {Shao}, {Shen}, {Small}, {Sohn}, {SooHoo}, {Tazaki}, {Tiede}, {Tilanus}, {Titus}, {Toma}, {Torne},
  {Trent}, {Trippe}, {Tsuda}, {van Bemmel}, {van Langevelde}, {van Rossum}, {Wagner}, {Wardle}, {Weintroub}, {Wex}, {Wharton}, {Wielgus}, {Wong}, {Wu}, {Young}, {Young}, {Younsi}, {Yuan}, {Yuan}, {Zensus}, {Zhao}, {Zhao}, {Zhu}, {Algaba}, {Allardi}, {Amestica}, {Anczarski}, {Bach}, {Baganoff}, {Beaudoin}, {Benson}, {Berthold}, {Blanchard}, {Blundell}, {Bustamente}, {Cappallo}, {Castillo-Dom{\'\i}nguez}, {Chang}, {Chang}, {Chang}, {Chen}, {Chilson}, {Chuter}, {C{\'o}rdova Rosado}, {Coulson}, {Crawford}, {Crowley}, {David}, {Derome}, {Dexter}, {Dornbusch}, {Dudevoir}, {Dzib}, {Eckart}, {Eckert}, {Erickson}, {Everett}, {Faber}, {Farah}, {Fath}, {Folkers}, {Forbes}, {Freund}, {G{\'o}mez-Ruiz}, {Gale}, {Gao}, {Geertsema}, {Graham}, {Greer}, {Grosslein}, {Gueth}, {Haggard}, {Halverson}, {Han}, {Han}, {Hao}, {Hasegawa}, {Henning}, {Hern{\'a}ndez-G{\'o}mez}, {Herrero-Illana}, {Heyminck}, {Hirota}, {Hoge}, {Huang}, {Impellizzeri}, {Jiang}, {Kamble}, {Keisler}, {Kimura}, {Kono}, {Kubo}, {Kuroda}, {Lacasse}, {Laing},
  {Leitch}, {Li}, {Lin}, {Liu}, {Liu}, {Lu}, {Marson}, {Martin-Cocher}, {Massingill}, {Matulonis}, {McColl}, {McWhirter}, {Messias}, {Meyer-Zhao}, {Michalik}, {Monta{\~n}a}, {Montgomerie}, {Mora-Klein}, {Muders}, {Nadolski}, {Navarro}, {Neilsen}, {Nguyen}, {Nishioka}, {Norton}, {Nowak}, {Nystrom}, {Ogawa}, {Oshiro}, {Oyama}, {Parsons}, {Paine}, {Pe{\~n}alver}, {Phillips}, {Poirier}, {Pradel}, {Primiani}, {Raffin}, {Rahlin}, {Reiland}, {Risacher}, {Ruiz}, {S{\'a}ez-Mada{\'\i}n}, {Sassella}, {Schellart}, {Shaw}, {Silva}, {Shiokawa}, {Smith}, {Snow}, {Souccar}, {Sousa}, {Sridharan}, {Srinivasan}, {Stahm}, {Stark}, {Story}, {Timmer}, {Vertatschitsch}, {Walther}, {Wei}, {Whitehorn}, {Whitney}, {Woody}, {Wouterloot}, {Wright}, {Yamaguchi}, {Yu}, {Zeballos}, {Zhang}, \& {Ziurys}}]{Akiyama2019}
{Event Horizon Telescope Collaboration}, {Akiyama}, K., {Alberdi}, A., {et~al.} 2019, \apjl, 875, L1, \dodoi{10.3847/2041-8213/ab0ec7}

\bibitem[{{Fabian}(2012)}]{Fabian2012}
{Fabian}, A.~C. 2012, \araa, 50, 455, \dodoi{10.1146/annurev-astro-081811-125521}

\bibitem[{{Farrah} {et~al.}(2023{\natexlab{a}}){Farrah}, {Petty}, {Croker}, {Tarl{\'e}}, {Zevin}, {Hatziminaoglou}, {Shankar}, {Wang}, {Clements}, {Efstathiou}, {Lacy}, {Nishimura}, {Afonso}, {Pearson}, \& {Pitchford}}]{Farrah2023a}
{Farrah}, D., {Petty}, S., {Croker}, K.~S., {et~al.} 2023{\natexlab{a}}, \apj, 943, 133, \dodoi{10.3847/1538-4357/acac2e}

\bibitem[{{Farrah} {et~al.}(2023{\natexlab{b}}){Farrah}, {Croker}, {Zevin}, {Tarl{\'e}}, {Faraoni}, {Petty}, {Afonso}, {Fernandez}, {Nishimura}, {Pearson}, {Wang}, {Clements}, {Efstathiou}, {Hatziminaoglou}, {Lacy}, {McPartland}, {Pitchford}, {Sakai}, \& {Weiner}}]{Farrah2023b}
{Farrah}, D., {Croker}, K.~S., {Zevin}, M., {et~al.} 2023{\natexlab{b}}, \apjl, 944, L31, \dodoi{10.3847/2041-8213/acb704}

\bibitem[{{French}(2021)}]{French2021}
{French}, K.~D. 2021, \pasp, 133, 072001, \dodoi{10.1088/1538-3873/ac0a59}

\bibitem[{{French} {et~al.}(2015){French}, {Yang}, {Zabludoff}, {Narayanan}, {Shirley}, {Walter}, {Smith}, \& {Tremonti}}]{French2015}
{French}, K.~D., {Yang}, Y., {Zabludoff}, A., {et~al.} 2015, \apj, 801, 1, \dodoi{10.1088/0004-637X/801/1/1}

\bibitem[{{Gebhardt} \& {Thomas}(2009)}]{Gebhardt2009}
{Gebhardt}, K., \& {Thomas}, J. 2009, \apj, 700, 1690, \dodoi{10.1088/0004-637X/700/2/1690}

\bibitem[{{Gim{\'e}nez-Arteaga} {et~al.}(2023){Gim{\'e}nez-Arteaga}, {Oesch}, {Brammer}, {Valentino}, {Mason}, {Weibel}, {Barrufet}, {Fujimoto}, {Heintz}, {Nelson}, {Strait}, {Suess}, \& {Gibson}}]{GimenezArteaga2023}
{Gim{\'e}nez-Arteaga}, C., {Oesch}, P.~A., {Brammer}, G.~B., {et~al.} 2023, \apj, 948, 126, \dodoi{10.3847/1538-4357/acc5ea}

\bibitem[{{G{\'o}mez} {et~al.}(2014){G{\'o}mez}, {Wyrowski}, {Schuller}, {Menten}, \& {Ballesteros-Paredes}}]{Gomez2014}
{G{\'o}mez}, L., {Wyrowski}, F., {Schuller}, F., {Menten}, K.~M., \& {Ballesteros-Paredes}, J. 2014, \aap, 561, A148, \dodoi{10.1051/0004-6361/201322310}

\bibitem[{{Greenberg} \& {Li}(1996)}]{Greenberg1996}
{Greenberg}, J.~M., \& {Li}, A. 1996, in Astrophysics and Space Science Library, Vol. 209, New Extragalactic Perspectives in the New South Africa, ed. D.~L. {Block} \& J.~M. {Greenberg}, 118, \dodoi{10.1007/978-94-009-0335-7_14}

\bibitem[{{Hallakoun} \& {Maoz}(2021)}]{Hallakoun2021}
{Hallakoun}, N., \& {Maoz}, D. 2021, \mnras, 507, 398, \dodoi{10.1093/mnras/stab2145}

\bibitem[{{Hammerstein} {et~al.}(2021){Hammerstein}, {Gezari}, {van Velzen}, {Cenko}, {Roth}, {Ward}, {Frederick}, {Hung}, {Graham}, {Foley}, {Bellm}, {Cannella}, {Drake}, {Kupfer}, {Laher}, {Mahabal}, {Masci}, {Riddle}, {Rojas-Bravo}, \& {Smith}}]{Hammerstein2021}
{Hammerstein}, E., {Gezari}, S., {van Velzen}, S., {et~al.} 2021, \apjl, 908, L20, \dodoi{10.3847/2041-8213/abdcb4}

\bibitem[{{H{\"a}ring} \& {Rix}(2004)}]{Haring2004}
{H{\"a}ring}, N., \& {Rix}, H.-W. 2004, \apjl, 604, L89, \dodoi{10.1086/383567}

\bibitem[{{Heintz} {et~al.}(2023){Heintz}, {Watson}, {Brammer}, {Vejlgaard}, {Hutter}, {Strait}, {Matthee}, {Oesch}, {Jakobsson}, {Tanvir}, {Laursen}, {Naidu}, {Mason}, {Killi}, {Jung}, {Hsiao}, {Abdurro'uf}, {Coe}, {Arrabal Haro}, {Finkelstein}, \& {Toft}}]{Heintz2023}
{Heintz}, K.~E., {Watson}, D., {Brammer}, G., {et~al.} 2023, arXiv e-prints, arXiv:2306.00647, \dodoi{10.48550/arXiv.2306.00647}

\bibitem[{{Heyer} \& {Dame}(2015)}]{Heyer2015}
{Heyer}, M., \& {Dame}, T.~M. 2015, \araa, 53, 583, \dodoi{10.1146/annurev-astro-082214-122324}

\bibitem[{{Hopkins}(2012)}]{Hopkins2012}
{Hopkins}, P.~F. 2012, \mnras, 423, 2037, \dodoi{10.1111/j.1365-2966.2012.20731.x}

\bibitem[{{Jeans}(1902)}]{Jeans1902}
{Jeans}, J.~H. 1902, Philosophical Transactions of the Royal Society of London Series A, 199, 1, \dodoi{10.1098/rsta.1902.0012}

\bibitem[{Jermyn {et~al.}(2018)Jermyn, Steinhardt, \& Tout}]{Jermyn2018}
Jermyn, A.~S., Steinhardt, C.~L., \& Tout, C.~A. 2018, Monthly Notices of the Royal Astronomical Society, 480, 4265–4272, \dodoi{10.1093/mnras/sty2123}

\bibitem[{{Kobayashi} \& {Nomoto}(2009)}]{Kobayashi2009}
{Kobayashi}, C., \& {Nomoto}, K. 2009, \apj, 707, 1466, \dodoi{10.1088/0004-637X/707/2/1466}

\bibitem[{{Kormendy} \& {Ho}(2013)}]{Kormendy2013}
{Kormendy}, J., \& {Ho}, L.~C. 2013, \araa, 51, 511, \dodoi{10.1146/annurev-astro-082708-101811}

\bibitem[{{Kormendy} \& {Richstone}(1995)}]{Kormendy1995}
{Kormendy}, J., \& {Richstone}, D. 1995, \araa, 33, 581, \dodoi{10.1146/annurev.aa.33.090195.003053}

\bibitem[{{Kroupa}(2001)}]{Kroupa2001}
{Kroupa}, P. 2001, \mnras, 322, 231, \dodoi{10.1046/j.1365-8711.2001.04022.x}

\bibitem[{{Kroupa}(2002)}]{Kroupa2002}
---. 2002, Science, 295, 82, \dodoi{10.1126/science.1067524}

\bibitem[{{Krumholz} \& {McKee}(2008)}]{Krumholz2008}
{Krumholz}, M.~R., \& {McKee}, C.~F. 2008, \nat, 451, 1082, \dodoi{10.1038/nature06620}

\bibitem[{{Labb{\'e}} {et~al.}(2023){Labb{\'e}}, {van Dokkum}, {Nelson}, {Bezanson}, {Suess}, {Leja}, {Brammer}, {Whitaker}, {Mathews}, {Stefanon}, \& {Wang}}]{Labbe2023}
{Labb{\'e}}, I., {van Dokkum}, P., {Nelson}, E., {et~al.} 2023, \nat, 616, 266, \dodoi{10.1038/s41586-023-05786-2}

\bibitem[{{Larson}(1985)}]{Larson1985}
{Larson}, R.~B. 1985, \mnras, 214, 379, \dodoi{10.1093/mnras/214.3.379}

\bibitem[{{Law-Smith} {et~al.}(2017){Law-Smith}, {Ramirez-Ruiz}, {Ellison}, \& {Foley}}]{LawSmith2017}
{Law-Smith}, J., {Ramirez-Ruiz}, E., {Ellison}, S.~L., \& {Foley}, R.~J. 2017, \apj, 850, 22, \dodoi{10.3847/1538-4357/aa94c7}

\bibitem[{{Li} {et~al.}(2018){Li}, {Mao}, {Emsellem}, {Xu}, {Springel}, \& {Krajnovi{\'c}}}]{Li2018}
{Li}, H., {Mao}, S., {Emsellem}, E., {et~al.} 2018, \mnras, 473, 1489, \dodoi{10.1093/mnras/stx2374}

\bibitem[{{Li} {et~al.}(2023){Li}, {Nair}, {Rowlands}, {Masters}, {Stark}, {Drory}, {Ellison}, {Irwin}, {Satyapal}, {Jones}, {Keel}, {Mukundan}, \& {Tu}}]{Li2023}
{Li}, W., {Nair}, P., {Rowlands}, K., {et~al.} 2023, \mnras, 523, 720, \dodoi{10.1093/mnras/stad1473}

\bibitem[{{Lopez} {et~al.}(2014){Lopez}, {Krumholz}, {Bolatto}, {Prochaska}, {Ramirez-Ruiz}, \& {Castro}}]{Lopez2014}
{Lopez}, L.~A., {Krumholz}, M.~R., {Bolatto}, A.~D., {et~al.} 2014, \apj, 795, 121, \dodoi{10.1088/0004-637X/795/2/121}

\bibitem[{{Low} \& {Lynden-Bell}(1976)}]{LyndenBell1976}
{Low}, C., \& {Lynden-Bell}, D. 1976, \mnras, 176, 367, \dodoi{10.1093/mnras/176.2.367}

\bibitem[{{Magnelli} {et~al.}(2014){Magnelli}, {Lutz}, {Saintonge}, {Berta}, {Santini}, {Symeonidis}, {Altieri}, {Andreani}, {Aussel}, {B{\'e}thermin}, {Bock}, {Bongiovanni}, {Cepa}, {Cimatti}, {Conley}, {Daddi}, {Elbaz}, {F{\"o}rster Schreiber}, {Genzel}, {Ivison}, {Le Floc'h}, {Magdis}, {Maiolino}, {Nordon}, {Oliver}, {Page}, {P{\'e}rez Garc{\'{\i}}a}, {Poglitsch}, {Popesso}, {Pozzi}, {Riguccini}, {Rodighiero}, {Rosario}, {Roseboom}, {Sanchez-Portal}, {Scott}, {Sturm}, {Tacconi}, {Valtchanov}, {Wang}, \& {Wuyts}}]{Magnelli2014}
{Magnelli}, B., {Lutz}, D., {Saintonge}, A., {et~al.} 2014, \aap, 561, A86, \dodoi{10.1051/0004-6361/201322217}

\bibitem[{{Magorrian} {et~al.}(1998){Magorrian}, {Tremaine}, {Richstone}, {Bender}, {Bower}, {Dressler}, {Faber}, {Gebhardt}, {Green}, {Grillmair}, {Kormendy}, \& {Lauer}}]{Magorrian1998}
{Magorrian}, J., {Tremaine}, S., {Richstone}, D., {et~al.} 1998, \aj, 115, 2285, \dodoi{10.1086/300353}

\bibitem[{{McConnell} \& {Ma}(2013)}]{McConnell2013}
{McConnell}, N.~J., \& {Ma}, C.-P. 2013, \apj, 764, 184, \dodoi{10.1088/0004-637X/764/2/184}

\bibitem[{{McLure} \& {Dunlop}(2004)}]{McLure2004}
{McLure}, R.~J., \& {Dunlop}, J.~S. 2004, \mnras, 352, 1390, \dodoi{10.1111/j.1365-2966.2004.08034.x}

\bibitem[{{Myeong} {et~al.}(2018){Myeong}, {Evans}, {Belokurov}, {Sanders}, \& {Koposov}}]{Myeong2018}
{Myeong}, G.~C., {Evans}, N.~W., {Belokurov}, V., {Sanders}, J.~L., \& {Koposov}, S.~E. 2018, \apjl, 863, L28, \dodoi{10.3847/2041-8213/aad7f7}

\bibitem[{Papadopoulos(2010)}]{Papadopoulos2010}
Papadopoulos, P.~P. 2010, ApJ, 720, 226.
\newblock \url{http://stacks.iop.org/0004-637X/720/i=1/a=226}

\bibitem[{{Parkash} {et~al.}(2018){Parkash}, {Brown}, {Jarrett}, \& {Bonne}}]{Parkash2018}
{Parkash}, V., {Brown}, M. J.~I., {Jarrett}, T.~H., \& {Bonne}, N.~J. 2018, \apj, 864, 40, \dodoi{10.3847/1538-4357/aad3b9}

\bibitem[{{Quintero} {et~al.}(2004){Quintero}, {Hogg}, {Blanton}, {Schlegel}, {Eisenstein}, {Gunn}, {Brinkmann}, {Fukugita}, {Glazebrook}, \& {Goto}}]{Quintero2004}
{Quintero}, A.~D., {Hogg}, D.~W., {Blanton}, M.~R., {et~al.} 2004, \apj, 602, 190, \dodoi{10.1086/380601}

\bibitem[{{Rowlands} {et~al.}(2018){Rowlands}, {Heckman}, {Wild}, {Zakamska}, {Rodriguez-Gomez}, {Barrera-Ballesteros}, {Lotz}, {Thilker}, {Andrews}, {Boquien}, {Brinkmann}, {Brownstein}, {Hwang}, \& {Smethurst}}]{Rowlands2018}
{Rowlands}, K., {Heckman}, T., {Wild}, V., {et~al.} 2018, \mnras, 480, 2544, \dodoi{10.1093/mnras/sty1916}

\bibitem[{{Rusakov} {et~al.}(2023){Rusakov}, {Steinhardt}, \& {Sneppen}}]{Rusakov2023}
{Rusakov}, V., {Steinhardt}, C.~L., \& {Sneppen}, A. 2023, \apjs, 268, 10, \dodoi{10.3847/1538-4365/acdde3}

\bibitem[{{Sahu} {et~al.}(2022){Sahu}, {Graham}, \& {Davis}}]{Sahu2022}
{Sahu}, N., {Graham}, A.~W., \& {Davis}, B.~L. 2022, \apj, 927, 67, \dodoi{10.3847/1538-4357/ac4251}

\bibitem[{{Salpeter}(1955)}]{Salpeter1955}
{Salpeter}, E.~E. 1955, \apj, 121, 161, \dodoi{10.1086/145971}

\bibitem[{{Shen} {et~al.}(2011){Shen}, {Richards}, {Strauss}, {Hall}, {Schneider}, {Snedden}, {Bizyaev}, {Brewington}, {Malanushenko}, {Malanushenko}, {Oravetz}, {Pan}, \& {Simmons}}]{Shen2011}
{Shen}, Y., {Richards}, G.~T., {Strauss}, M.~A., {et~al.} 2011, \apjs, 194, 45, \dodoi{10.1088/0067-0049/194/2/45}

\bibitem[{{Shen} {et~al.}(2024){Shen}, {Grier}, {Horne}, {Stone}, {Li}, {Yang}, {Homayouni}, {Trump}, {Anderson}, {Brandt}, {Hall}, {Ho}, {Jiang}, {Petitjean}, {Schneider}, {Tao}, {Donnan}, {AlSayyad}, {Bershady}, {Blanton}, {Bizyaev}, {Bundy}, {Chen}, {Davis}, {Dawson}, {Fan}, {Greene}, {Gr{\"o}ller}, {Guo}, {Ibarra-Medel}, {Jiang}, {Keenan}, {Kollmeier}, {Lejoly}, {Li}, {de la Macorra}, {Moe}, {Nie}, {Rossi}, {Smith}, {Tee}, {Weijmans}, {Xu}, {Yue}, {Zhou}, {Zhou}, \& {Zou}}]{Shen2024}
{Shen}, Y., {Grier}, C.~J., {Horne}, K., {et~al.} 2024, \apjs, 272, 26, \dodoi{10.3847/1538-4365/ad3936}

\bibitem[{{Sneppen} {et~al.}(2022){Sneppen}, {Steinhardt}, {Hensley}, {Jermyn}, {Mostafa}, \& {Weaver}}]{Sneppen2022}
{Sneppen}, A., {Steinhardt}, C.~L., {Hensley}, H., {et~al.} 2022, \apj, 931, 57, \dodoi{10.3847/1538-4357/ac695e}

\bibitem[{{Sommovigo} {et~al.}(2022){Sommovigo}, {Ferrara}, {Pallottini}, {Dayal}, {Bouwens}, {Smit}, {da Cunha}, {De Looze}, {Bowler}, {Hodge}, {Inami}, {Oesch}, {Endsley}, {Gonzalez}, {Schouws}, {Stark}, {Stefanon}, {Aravena}, {Graziani}, {Riechers}, {Schneider}, {van der Werf}, {Algera}, {Barrufet}, {Fudamoto}, {Hygate}, {Labb{\'e}}, {Li}, {Nanayakkara}, \& {Topping}}]{Sommovigo2022}
{Sommovigo}, L., {Ferrara}, A., {Pallottini}, A., {et~al.} 2022, \mnras, 513, 3122, \dodoi{10.1093/mnras/stac302}

\bibitem[{{Spinrad}(1973)}]{Spinrad1973}
{Spinrad}, H. 1973, \apj, 182, 381, \dodoi{10.1086/152146}

\bibitem[{{Steinhardt} {et~al.}(2016){Steinhardt}, {Capak}, {Masters}, \& {Speagle}}]{Steinhardt2016}
{Steinhardt}, C.~L., {Capak}, P., {Masters}, D., \& {Speagle}, J.~S. 2016, \apj, 824, 21, \dodoi{10.3847/0004-637X/824/1/21}

\bibitem[{{Steinhardt} \& {Elvis}(2010)}]{Steinhardt2010}
{Steinhardt}, C.~L., \& {Elvis}, M. 2010, \mnras, 402, 2637, \dodoi{10.1111/j.1365-2966.2009.16084.x}

\bibitem[{{Steinhardt} \& {Elvis}(2011)}]{Steinhardt2011}
---. 2011, \mnras, 410, 201, \dodoi{10.1111/j.1365-2966.2010.17435.x}

\bibitem[{{Steinhardt} {et~al.}(2022{\natexlab{a}}){Steinhardt}, {Kokorev}, {Rusakov}, {Garcia}, \& {Sneppen}}]{Steinhardt2023}
{Steinhardt}, C.~L., {Kokorev}, V., {Rusakov}, V., {Garcia}, E., \& {Sneppen}, A. 2022{\natexlab{a}}, arXiv e-prints, arXiv:2208.07879.
\newblock \doarXiv{2208.07879}

\bibitem[{{Steinhardt} \& {Speagle}(2014)}]{Steinhardt2014b}
{Steinhardt}, C.~L., \& {Speagle}, J.~S. 2014, \apj, 796, 25, \dodoi{10.1088/0004-637X/796/1/25}

\bibitem[{{Steinhardt} {et~al.}(2022{\natexlab{b}}){Steinhardt}, {Sneppen}, {Mostafa}, {Hensley}, {Jermyn}, {Lopez}, {Weaver}, {Brammer}, {Clark}, {Davidzon}, {Diaconu}, {Mobasher}, {Rusakov}, \& {Toft}}]{Steinhardt2022a}
{Steinhardt}, C.~L., {Sneppen}, A., {Mostafa}, B., {et~al.} 2022{\natexlab{b}}, \apj, 931, 58, \dodoi{10.3847/1538-4357/ac62d6}

\bibitem[{{Steinhardt} {et~al.}(2022{\natexlab{c}}){Steinhardt}, {Sneppen}, {Hensley}, {Jermyn}, {Mostafa}, {Weaver}, {Brammer}, {Clark}, {Davidzon}, {Diaconu}, {Mobasher}, {Rusakov}, \& {Toft}}]{Steinhardt2022b}
{Steinhardt}, C.~L., {Sneppen}, A., {Hensley}, H., {et~al.} 2022{\natexlab{c}}, arXiv e-prints, arXiv:2206.01750.
\newblock \doarXiv{2206.01750}

\bibitem[{{Vestergaard} \& {Peterson}(2006)}]{Vestergaard2006}
{Vestergaard}, M., \& {Peterson}, B.~M. 2006, \apj, 641, 689, \dodoi{10.1086/500572}

\bibitem[{{Weaver} {et~al.}(2022){Weaver}, {Kauffmann}, {Ilbert}, {McCracken}, {Moneti}, {Toft}, {Brammer}, {Shuntov}, {Davidzon}, {Hsieh}, {Laigle}, {Anastasiou}, {Jespersen}, {Vinther}, {Capak}, {Casey}, {McPartland}, {Milvang-Jensen}, {Mobasher}, {Sanders}, {Zalesky}, {Arnouts}, {Aussel}, {Dunlop}, {Faisst}, {Franx}, {Furtak}, {Fynbo}, {Gould}, {Greve}, {Gwyn}, {Kartaltepe}, {Kashino}, {Koekemoer}, {Kokorev}, {Le F{\`e}vre}, {Lilly}, {Masters}, {Magdis}, {Mehta}, {Peng}, {Riechers}, {Salvato}, {Sawicki}, {Scarlata}, {Scoville}, {Shirley}, {Silverman}, {Sneppen}, {Smolc̆i{\'c}}, {Steinhardt}, {Stern}, {Tanaka}, {Taniguchi}, {Teplitz}, {Vaccari}, {Wang}, \& {Zamorani}}]{Weaver2022}
{Weaver}, J.~R., {Kauffmann}, O.~B., {Ilbert}, O., {et~al.} 2022, \apjs, 258, 11, \dodoi{10.3847/1538-4365/ac3078}

\bibitem[{{Wild} {et~al.}(2009){Wild}, {Walcher}, {Johansson}, {Tresse}, {Charlot}, {Pollo}, {Le F{\`e}vre}, \& {de Ravel}}]{Wild2009}
{Wild}, V., {Walcher}, C.~J., {Johansson}, P.~H., {et~al.} 2009, \mnras, 395, 144, \dodoi{10.1111/j.1365-2966.2009.14537.x}

\bibitem[{{Williams} {et~al.}(2009){Williams}, {Quadri}, {Franx}, {van Dokkum}, \& {Labb{\'e}}}]{Williams2009}
{Williams}, R.~J., {Quadri}, R.~F., {Franx}, M., {van Dokkum}, P., \& {Labb{\'e}}, I. 2009, \apj, 691, 1879, \dodoi{10.1088/0004-637X/691/2/1879}

\bibitem[{{Worthey} \& {Ottaviani}(1997)}]{Worthey1997}
{Worthey}, G., \& {Ottaviani}, D.~L. 1997, \apjs, 111, 377, \dodoi{10.1086/313021}

\bibitem[{{Zabludoff} {et~al.}(1996){Zabludoff}, {Zaritsky}, {Lin}, {Tucker}, {Hashimoto}, {Shectman}, {Oemler}, \& {Kirshner}}]{Zabludoff1996}
{Zabludoff}, A.~I., {Zaritsky}, D., {Lin}, H., {et~al.} 1996, \apj, 466, 104, \dodoi{10.1086/177495}

\bibitem[{{Zhang} {et~al.}(2023){Zhang}, {Li}, {Leja}, {Whitaker}, {Nersesian}, {Bezanson}, \& {van der Wel}}]{Zhang2023}
{Zhang}, J., {Li}, Y., {Leja}, J., {et~al.} 2023, \apj, 952, 6, \dodoi{10.3847/1538-4357/acd84a}

\end{thebibliography}



\label{lastpage}
\end{document}